\documentclass{article}

\usepackage{amsmath,amssymb,amsfonts}
\usepackage{microtype}
\usepackage{graphicx}
\usepackage{subfigure}
\usepackage{booktabs} 

\DeclareMathOperator*{\argmax}{arg\,max}

\usepackage{hyperref}

\usepackage[accepted]{icml2020}

\icmltitlerunning{Graph Neural Networks for Massive MIMO Detection}

\begin{document}

\twocolumn[
\icmltitle{Graph Neural Networks for Massive MIMO Detection}




\begin{icmlauthorlist}
\icmlauthor{Andrea Scotti}{kth,huawei}
\icmlauthor{Nima N. Moghadam}{huawei}
\icmlauthor{Dong Liu}{kth}
\icmlauthor{Karl Gafvert}{huawei}
\icmlauthor{Jinliang Huang}{huawei}
\end{icmlauthorlist}

\icmlaffiliation{kth}{KTH Royal Institute of Technology, Stockholm, Sweden}
\icmlaffiliation{huawei}{Huawei Technologies Sweden AB, Stockholm, Sweden}

\icmlcorrespondingauthor{Andrea Scotti}{scotti@kth.se}

\icmlkeywords{Machine Learning, ICML}

\vskip 0.3in
]



\printAffiliationsAndNotice{}  

\begin{abstract}
In this paper, we innovately use graph neural networks (GNNs) to learn a message-passing solution for the inference task of massive multiple multiple-input multiple-output (MIMO) detection in wireless communication. We adopt a graphical model based on the Markov random field (MRF) where belief propagation (BP) yields poor results when it assumes a uniform prior over the transmitted symbols. Numerical simulations show that, under the uniform prior assumption, our GNN-based MIMO detection solution outperforms the minimum mean-squared error (MMSE) baseline detector, in contrast to BP. Furthermore, experiments demonstrate that the performance of the algorithm slightly improves by incorporating MMSE information into the prior.
\end{abstract}

\section{Introduction}
Massive MIMO (multiple-input and multiple-output) is a method to improve the spectral efficiency and link reliability of wireless communication systems \cite{1203154}, by having a large number of transmitter and receiver antennas. In the fifth-generation (5G) mobile communication system, massive MIMO is a key technology to face the increasing number of mobile users and satisfy user demands. One of the challenging problems in massive MIMO is to design efficient detection algorithms for recovering the transmitted information from multiple users \cite{8804165}. 
The optimal solution for the MIMO detection problem is the maximum likelihood (ML) detector \cite{8804165}. However, ML detection is not used in practice because its complexity  increases exponentially with the number of transmitters. A number of sub-optimal solutions have been proposed to balance the trade-off between performance and complexity, e.g., sphere decoding (SD) \cite{guo2006algorithm}, zero-forcing (ZF) and minimum mean-squared error (MMSE) detectors \cite{mmse}, etc. In the last decade, methods based on probabilistic graphical models (PGM) have been actively studied \cite{gta, goldberger2010pseudoprior, liu2019}, where the MIMO detection problem is firstly modeled by a maximum a posteriori (MAP) inference task in a pairwise Markov random field (MRF) and then addressed approximately with belief propagation (BP) \cite{Yedidia2003UnderstandingBP}. BP is an iterative message-passing algorithm for performing exact inference on tree-structured graphical models. Its low complexity and efficiency, even for general graphs, make it very attractive for massive MIMO detection. However, due to dense connections in the MRF graph representation of the MIMO problem, BP's performance is sensitive to both prior information and the message update rules.

In this work, we innovatively use graph neural networks (GNNs) to learn a message-passing solution that addresses the inference task of MIMO detection. Specifically, our approach is built upon the GNN framework developed in \cite{yoon2018inference}. Instead of propagating messages by hand-crafted update functions as in BP, \cite{yoon2018inference} uses neural networks to learn the message-passing rules and give approximate updates.

Our network is called MIMO-GNN and it can solve MIMO detection under time-varying channels and higher-order qadrature amplitude modulation (QAM), such as 16-QAM. In practice, the correlation in the channel is not known a priori. Therefore, MIMO-GNN is trained on independent and identically distributed (i.i.d.) and Gaussian distributed channels and then tested on correlated channels drawn from a different distribution (specifically, the Kronecker model \cite{kronecker}).

\textit{Notations--} We denote the transpose, the \((i,j)\) entry and the \(j\)-th column of maxtrix \(\mathbf{A}\), by \(\mathbf{A}^T\), \(a_{ij}\) and \(\mathbf{a}_j\), respectively. \(a_i\) stands for the \(i\)-th entry of vector \(\mathbf{a}\). \(\mathbf{I}_N\) denotes the identity matrix of shape \(N\).


\section{Background}\label{background}

\subsection{MIMO Detection}
Wireless communication in a MIMO system requires coordination of multiple antennas at the receiver unit to detect the signals sent from wireless devices. These devices operate as mobile transmitters within a limited coverage area commonly known as cell. Here, we consider the uplink communication in a cellular system where the base station has the role of a central coordinator. The MIMO system described above can be modeled by the real-valued linear system
\begin{equation} \label{mimo_equation}
	\mathbf{y} = \mathbf{H}\mathbf{x}+\mathbf{n}.
\end{equation}
The goal of MIMO detection is to infer the transmitted signal vector \({\mathbf{x}} \in {\mathcal{A}}^{{N}_t}\) where \({\mathcal{A}} \subset \mathbb{R}\) is a discrete finite alphabet (\(|\mathcal{A}|=\sqrt{M}\) according to \(M\)-QAM) and \({N}_t\) is the number of transmitted symbols.
The channel matrix \({\mathbf{H}} \in \mathbb{R}^{{N}_r} \times \mathbb{R}^{{N}_t}\) and measurement vector \({\mathbf{y}} \in \mathbb{R}^{{N}_r}\) are known variables where \({N}_r\) is the number of received symbols. The noise vector \({\mathbf{n}} \in \mathbb{R}^{{N}_r}\) is zero-mean Gaussian \({\mathbf{n}} \sim \mathcal{N}(0, \sigma^2\mathbf{I}_{{N}_r})\). The real-valued system described above is derived from the actual complex-valued system where the number of receiver and transmitter antennas are \(\frac{N_r}{2}\) and \(\frac{N_t}{2}\) respectively. More details regarding the conversion from the complex-valued to the real-valued system is provided in \cite{gta}.
	 
\subsection{Pair-wise MRF} \label{pair_wise_MRF}
A MRF models the structured dependency of a set of random variables \(\mathbf{x}=\{x_0, ..., x_{N-1}\}\) by an undirected graph \(\mathcal{G} = \{V,E\}\), where \(V\) and \(E\) are the set of nodes and edges respectively. Every node \(i \in V\) is associated to variable \(x_i\) and it holds that \(p(x_i|\mathbf{x} \backslash x_i) = p(x_i|ne(i))\), where $\backslash$ denotes exclusion and \(ne(i)\) is the set of neighbors of node \(i\).
In a pair-wise MRF a self potential \(\phi_i(x_i)\) is assigned to node \(i \in V\) and a pair potential \(\phi_{ij}(x_i, x_j)\) is assigned to the edge \(e \in E\) that connects node \(i \in V\) to node \(j \in V\).
The probability distribution corresponding to a pair-wise MRF has the following form:
\begin{equation}\label{factorized_prob}
	p(\mathbf{x}) = \frac{1}{Z}\prod_{i \in V}\phi_i(x_i)\prod_{(i,j)\in E}\phi_{ij}(x_i, x_j),
\end{equation}
where \(Z\) is a normalization constant.\\
In order to obtain an approximation \(b_l(x_l)\) of the marginal distribution for the variable \(x_l\), we can run the iterative message-passing algorithm, BP \cite{Yedidia2003UnderstandingBP}.

\begin{figure}[!t]
	\begin{center}
  		\includegraphics[scale=0.35]{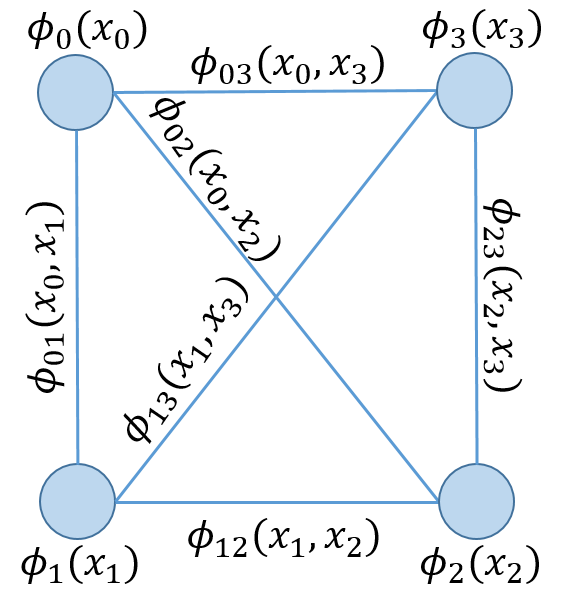}
  		\vskip -0.17in
  	    \caption{Fully-connected pair-wise Markov Random Fields with 4 variables.}
  \label{gnn_messages}
  \end{center}
  \vskip -0.25in
\end{figure}

\subsection{MIMO as a Markov Random Field} \label{MIMO_MRF}
Given the constrained linear system in \eqref{mimo_equation}, the corresponding posterior probability \(p(\mathbf{x}|\mathbf{y})\) is factorized according to the Bayes's rule in the following way: 
\begin{equation}\label{posteriori}
	p(\mathbf{x}|\mathbf{y}) \propto p(\mathbf{y}|\mathbf{x})p(\mathbf{x}) = \exp\left\{-\frac{1}{2\sigma^2}||\mathbf{Hx}-\mathbf{y}||^2\right\}p(\mathbf{x}),
\end{equation}
where \(p(\mathbf{x})\) is the prior distribution for \(\mathbf{x}\).
The goal of MIMO detection is to solve the following MAP problem:
\begin{equation}\label{MAP}
	\hat{\mathbf{x}}_{MAP} = \argmax_{\mathbf{x} \in \mathcal{A}^{Nt}}p(\mathbf{x}|\mathbf{y}).
\end{equation}
The posterior probability in \eqref{posteriori} can be factorized into a pair-wise MRF as in \eqref{factorized_prob} by assignment
\begin{align}
	\phi_i(x_i) &= e^{\frac{1}{\sigma^2}(\mathbf{y}^T\mathbf{h}_i x_i-\frac{1}{2}\mathbf{h}_i^T\mathbf{h}_ix_i^2)}p_i(x_i), \\
	\phi_{ij}(x_i, x_j) &= e^{-\frac{1}{\sigma^2}\mathbf{h}_i^T\mathbf{h}_jx_ix_j},
\end{align}
where \(\sigma^2\) is the noise variance and \(\mathbf{h}_i\) is the \(i\)-column of \(\mathbf{H}\).
By applying the BP algorithm, where the initial messages are the uniform prior probabilities over symbols, we can approximate the solution of the MAP problem in Equation \eqref{MAP} by solving a simplified MAP problem for each variable. Indeed, after convergence of BP, for each variable \(x_l\) we compute the belief \(b_l(x_l)\) as a function of the updated messages \cite{Yedidia2003UnderstandingBP} and hard detect the transmitted symbol \(\hat{x}_l\) with
\begin{equation}
	\hat{x}_l = \argmax_{x_l \in \mathcal{A}}b_l(x_l), ~~\forall~~l.
\end{equation}

\subsection{GNNs}\label{gnn_section}
GNNs from \cite{yoon2018inference} combine the advantages of deep learning and MRFs in a unique framework to capture the structure of the data into feature vectors that are updated through message-passing between nodes. In a GNN, a vector \(\mathbf{u}_i \in \mathbb{R}^{S_u}\), where \(S_u\) is a positive integer, encodes the information of a variable node in a MRF \eqref{factorized_prob}. The values of $\left\{\mathbf{u}_i\right\}$ are iteratively updated by a recurrent neural network (RNN) with input including the value of \(\mathbf{u}_i\) at the previous iteration together with the information coming from the neighbor nodes states \(\mathbf{u}_j : j \in ne(i)\) on the specified graph \(\mathcal{G}\) defined in Section \ref{pair_wise_MRF}.

The network is composed by three main modules: a propagation, an aggregation and a readout module. The first  two modules operate at every iteration \(t\) while the readout module is involved only after the last iteration \(T\). The propagation module outputs the updated message \(\mathbf{m}_{i \rightarrow j}^{t}\) for each direct edge \(e_{ij} \in E\)
\begin{equation}
	\mathbf{m}_{i \rightarrow j}^{t} = M(\mathbf{u}_i^{t-1}, \mathbf{u}_j^{t-1}, \epsilon_{ij}),
\end{equation}
where \(\epsilon_{ij}\) is the information associated to the edge \(e_{ij}\) and \(M\) is a multiple layer perceptron (MLP) with ReLU as activation functions. Therefore, the information exchanged between two nodes at iteration \(t\) is an encoding of the concatenation of the feature vectors of the two nodes and the information along the direct edge between them. The aggregation module operates at a node level by aggregating the incoming messages \(\mathbf{m}_{j \rightarrow i}^{t}\) at node \(i \in V\), with \(j \in ne(i)\), by following
\begin{equation}
	\mathbf{u}_i^{t} = U(\mathbf{u}_{i}^{t-1}, \sum_{j \in ne(i)}\mathbf{m}_{j \rightarrow i}^{t}),
\end{equation}
where \(U\) is a GRU \cite{Cho_2014}.

After \(T\) iterations, the feature vectors \(\mathbf{u}_i\) are used to make inference with the readout module. If the problem that we want to solve is to compute the marginal probabilities of discrete random variables, the readout module is a MLP \(R\) of the feature vector \(\mathbf{u}_i\) followed by the softmax function \(\gamma: \mathbb{R}^{|\mathcal{A}|} \rightarrow \mathbb{R}^{|\mathcal{A}|}\). The softmax function maps the non-normalized output \(\mathbf{z}\) of the network \(R\) to a probability distribution over predicted output symbols \(s_k\)
\begin{equation}
	\hat{p}(x_i = s_k) = \gamma(\mathbf{z})_k = \frac{e^{z_k}}{\sum_{j=1}^{|\mathcal{A}|}e^{z_j}}.
\end{equation}
The parameters of \(M, U\) and \(R\) are shared across the whole graph and we learn them with supervised learning by minimizing the loss function between the true probabilities \(p(x_i)\) and the predicted ones \(\hat{p}(x_i)\). A good candidate for the loss function \(L\) is the cross-entropy:
\begin{equation} \label{cross-entropy}
	L = -\sum_{x_i}p(x_i)\log\hat{p}(x_i),
\end{equation}
where \(\hat{p}(x_i)\) is the output of the \(T\)-layers GNN.

\begin{figure}[!t]
	\begin{center}
  		\includegraphics[width=\columnwidth]{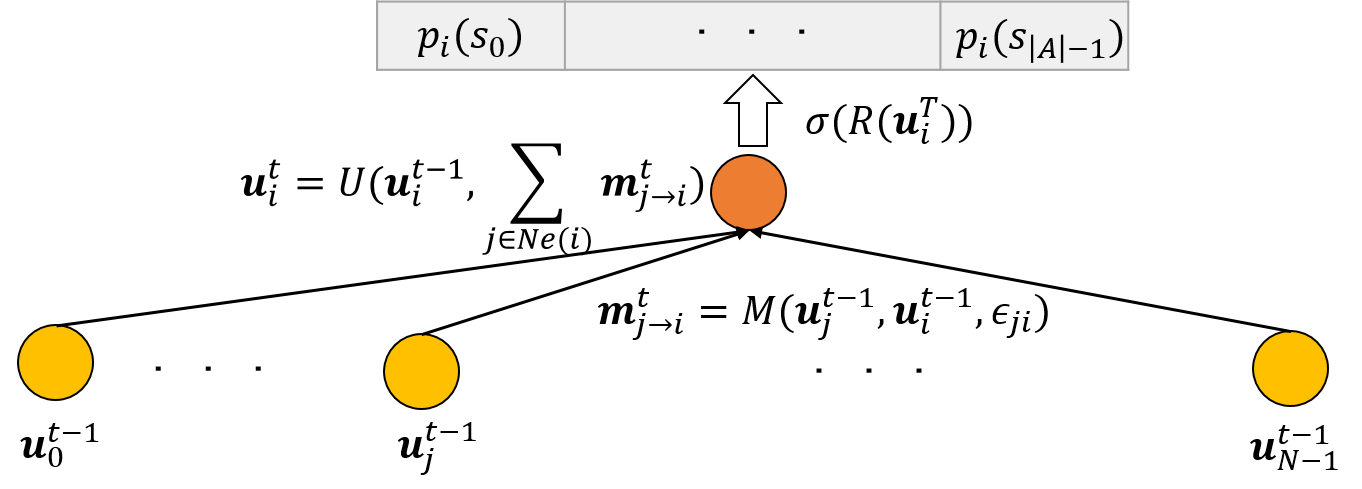}
  		\vskip -0.17in
  		\caption{Message, state and output updates in GNNs.}
  \label{gnn_messages}
  \end{center}
  \vskip -0.25in
\end{figure}
\section{Algorithms Design}\label{algorithms_design}
\subsection{MIMO-GNN} \label{GNN_noprior}
The GNN framework presented in Section \ref{gnn_section} can be used to infer the a posteriori probability \(p(\mathbf{x}|\mathbf{y})\) and recover the transmitted symbols \(\mathbf{x}\) in the MIMO detection problem in \eqref{mimo_equation}.
In this case, GNNs are built upon the MIMO MRF presented in Section \ref{MIMO_MRF}. Indeed, the input of GNNs is extracted from \(\phi_i(x_i)\) and \(\phi_{ij}(x_i, x_j)\). The information \(\epsilon_{ij}\) along each edge \(e_{ij}\) is the feature vector \(
	\epsilon_{ij} = [-\mathbf{h}_i^T\mathbf{h}_j, \;\sigma^2]\). The hidden vector \(\mathbf{u}_i^{0}\) of each node \(i\) is initialized with \(\mathbf{u}_i^{0} = \mathbf{W}[\mathbf{y}^T\mathbf{h}_i,\; \mathbf{h}_i^T\mathbf{h}_i, \; \sigma^2]^T+\mathbf{b}\).
Since we want to work with an hidden state of a given size \(S_u\), to simplify the implementation we encode the initial vector \([\mathbf{y}^T\mathbf{h}_i,\; \mathbf{h}_i^T\mathbf{h}_i, \; \sigma^2]\) with a linear transformation given by a learnable matrix \( \mathbf{W} \in \mathbb{R}^{S_u} \times \mathbb{R}^3\) and a learnable vector \(\mathbf{b} \in \mathbb{R}^{S_u}\).
The functions \(M\) and \(R\) are two different neural networks with two hidden layers and ReLU as activation functions. Both \(M\) and \(R\) implement dropout between hidden layers (rate of 0.1 in \(M\) and rate of 0.2 in \(R\)). The outputs of the first and the second hidden layer have sizes \(l\) and \(\frac{l}{2}\) respectively. Instead, the function \(U\) is composed of a GRU network followed by a linear layer that ensures that the output size is equal to \(S_u\). In the experiments the dimension of the GRU hidden state is \(l\).

Since in modern wireless communication systems soft symbols are more suitable than predicted symbols without probabilistic information, the predicted value \(\hat{x}_l\) for the transmitted symbol is the expected value of \(x_l\) with probability distribution \(\hat{p}(x_l)\):
\begin{equation}
	\hat{x}_l = \mathbb{E}_{x_l}\{x_l\} = \sum_{s \in \mathcal{A}}s\hat{p}(s).
\end{equation}
Similarly to BP for the fully connected pair-wise MRF, the complexity of MIMO-GNN is proportional to the number of edges in every iteration. However, for each edge, we need to perform a forward step in a feed-forward neural network, which increases the overall complexity.

\subsection{MIMO-GNN-MMSE} \label{GNN-prior}
In the previous section, we solve MIMO detection by assuming a uniform prior \(p(\mathbf{x})\) over the unknown symbols \(\mathbf{x}\). In this section, to improve the prior information, we incorporate the MMSE posterior as the prior \(p(\mathbf{x})\) such that
\begin{equation}
	p_l(x_l) = \frac{1}{\sqrt{2\pi c_{ll}}}\exp{(-\frac{(z_l-x_l)^2}{2c_{ll}})},
\end{equation}
where  $z_l$ is the $l$-th element of the MMSE estimation vector \(\mathbf{z}= (\mathbf{H}^T\mathbf{H}+\sigma^2\mathbf{I}_{N_t})^{-1}\mathbf{H}^T\mathbf{y} \label{z_mmse}\) and $c_{ll}$ is the $(l,l)$ element of \(\mathbf{C} = \sigma^2(\mathbf{H}^T\mathbf{H}+\sigma^2\mathbf{I}_{N_t})^{-1} \label{C_mmse}\).
The prior correlation coefficient \(\rho_{ij}\) between the variable \(x_i\) and \(x_j\), \(\rho_{ij} = \frac{c_{ij}^2}{c_{ii}c_{jj}}\), is added to the feature vector \(\epsilon_{ij}\).

In the implementation we reuse the same model in Section \ref{GNN_noprior} (with the same hyperparameters) and we only modify the information \(\epsilon_{ij}\) along the edges and the initial value of the hidden states \(\mathbf{u}_i^{0}\). The information along each edge \(e_{ij}\) becomes \(
	\epsilon_{ij} = [\rho_{ij}, \; -\mathbf{h}_i^T\mathbf{h}_j, \;\sigma^2]\), and the initial hidden vector \(\mathbf{u}_i^{0}\) of each node \(i\) is initialized with \(\mathbf{u}_i^{0} = \mathbf{W}[z_i, \; c_{ii}, \; \mathbf{y}^T\mathbf{h}_i,\; \mathbf{h}_i^T\mathbf{h}_i, \; \sigma^2]^T+\mathbf{b}\), where \( \mathbf{W} \in \mathbb{R}^{S_u} \times \mathbb{R}^5\) is a learnable matrix and \(\mathbf{b} \in \mathbb{R}^{S_u}\) is a learnable vector.
	
	MIMO-GNN-MMSE exhibits a higher complexity than MIMO-GNN due to the computation of \(\mathbf{z}\) and \(\mathbf{C}\) that require the inversion of a matrix of size \(N_t \times N_t\).
	
\section{Numerical Experiments}\label{dataset_training}

We consider a MIMO configuration with 16 transmitter antennas (\(N_t = 32\)) and 32 receiver antennas (\(N_r = 64\)). The modulation scheme is 16-QAM. 

To synthetically build the datasets (for training, validation and testing) we use three sources of randomness in each sample: signal \(\mathbf{x}\), channel noise \(\mathbf{n}\) and channel matrix \(\mathbf{H}\). We ensure that the transmitter power satisfies \(\mathbb{E}\{\mathbf{x}^T\mathbf{x}\} = N_t\). The transmitted signal \(\mathbf{x}\) is generated randomly and uniformly over the corresponding constellation set. The channel noise standard deviation \(\sigma\) is derived from the definition of SNR: \[\text{SNR} = 10\log_{10}\frac{\mathbb{E}\{||\mathbf{H}\mathbf{x}||_2^2\}}{\mathbb{E}\{||\mathbf{n}||_2^2\}} = 10\log_{10}\frac{\mathbb{E}\{||\mathbf{H}\mathbf{x}||_2^2\}}{N_r \sigma^2}.\]
MIMO-GNN and MIMO-GNN-MMSE are both trained on a pre-built dataset of size 65536 and batch size 64.  The size of the (additional) validation dataset is 25\% of the training dataset size. The noise standard deviation \(\sigma\) is fixed within each batch. Since the dataset labels must be a discrete probability distribution \(p(s)\) over the constellation symbols \(s \in \mathcal{A}^{N_t}\), we opt for one-hot encoded labels where \(p(s)=1\) when \(s=x_l\) and 0 otherwise, where \(x_l\) is the transmitted symbol.

MIMO-GNN and MIMO-GNN-MMSE are both trained with early stopping, Adam optimizer and a learning rate of 0.0001 to minimize the loss \(L\) defined in \eqref{cross-entropy}. Since the correlation in the channel is not known a priori, the training is performed over channel matrices randomly sampled from the i.i.d. Gaussian channel model, where it holds that \(h_{ij} \sim \mathcal{N}(0, \frac{1}{N_r})\) for each element of \(\mathbf{H}\). After cross-validation, the hyperparameters are chosen to be \(l=128, \; S_m=S_u=8, \; T=10\). 

MMSE, BP, MIMO-GNN and MIMO-GNN-MMSE are tested on the Kronecker channel model that controls the correlation in the MIMO channel through a correlation coefficient \(\rho\), according to the exponential correlation model \cite{kronecker} that structures the channel matrix \(\mathbf{H}\) as follows: \(\mathbf{H} = \mathbf{R}_R^{1/2}\mathbf{K}\mathbf{R}_T^{1/2}.\)
Here, \(k_{ij} \sim \mathcal{N}(0,\frac{1}{N_r})\) and \(\mathbf{R}_R, \mathbf{R}_T\) are the spatial correlation matrices at the receiver and the transmitter side respectively.

The performances of the algorithms are tested according to the symbol error rate (SER) metric. 
The results are averaged over a (additional) dataset of 20000 random simulations. 
BP runs for 8 iterations and implements a damping factor of 0.75 on belief and messages \cite{5503188} to increase the performance. Moreover, the prior \(p_l(x_l)\) at iteration \(t+1\) is improved with the belief \(b_l(x_l)\) computed at iteration \(t\).

\begin{figure}[!t]
	\begin{center}
  		\includegraphics[width=\columnwidth]{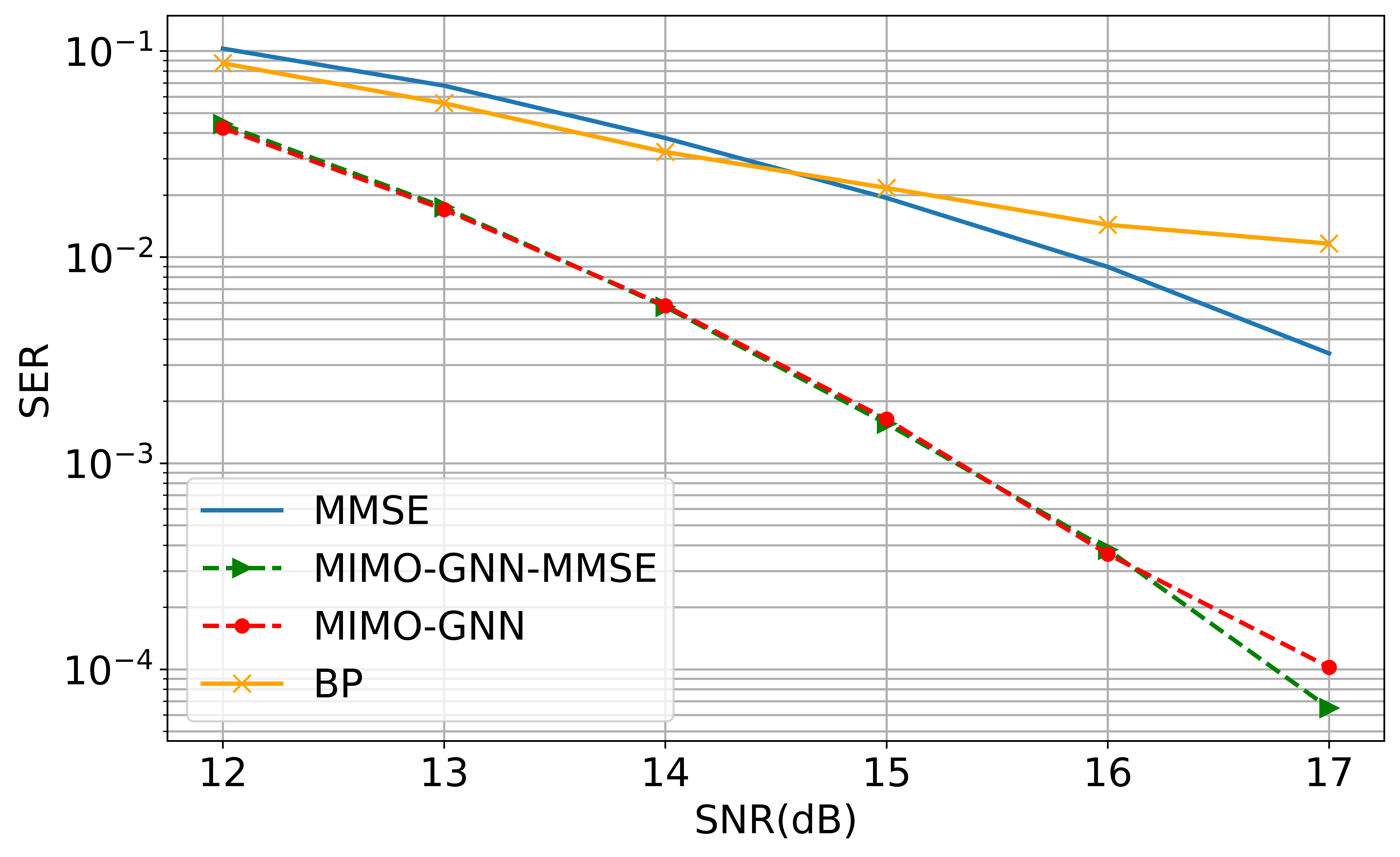}
  		\vskip -0.19in
  		\caption{SER vs. SNR of different schemes for 16-QAM modulation, MIMO system with \(N_t=32\) and \(N_r=64\) and channels i.i.d. and Gaussian distributed.}
  		\label{iid}
  \end{center}
  \vskip -0.13in
\end{figure}

\begin{figure}[!t]
	\begin{center}
  		\includegraphics[width=\columnwidth]{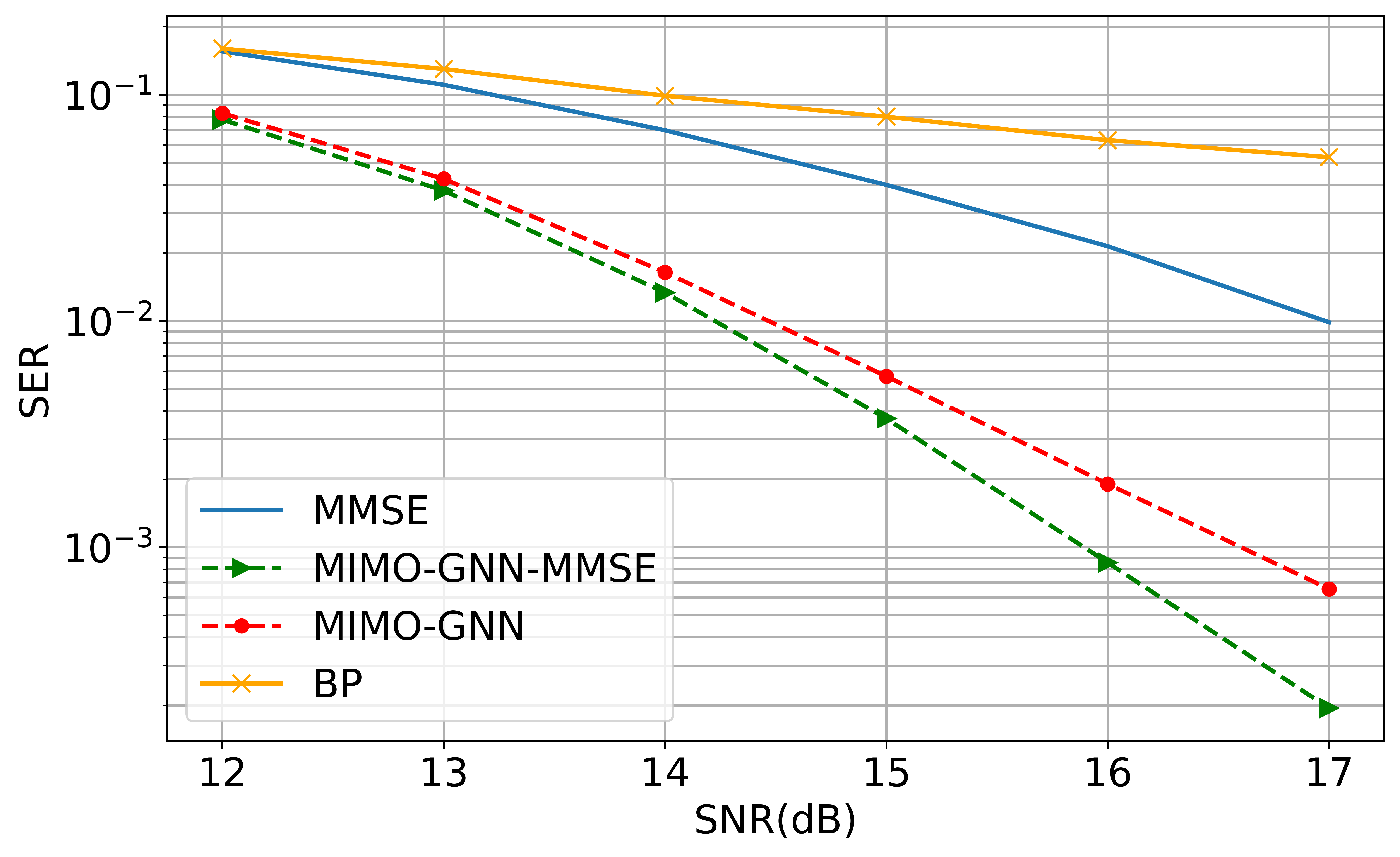}
  		\vskip -0.19in
  		\caption{SER vs. SNR of different schemes for 16-QAM modulation, MIMO system with \(N_t=32\) and \(N_r=64\) and channels randomly sampled from Kronecker model with \(\rho=0.3\).}
  		\label{kro}
  \end{center}
  \vskip -0.25in
\end{figure}

Fig. \ref{iid} shows the results for i.i.d. and Gaussian distributed channels (Kronecker model with \(\rho=0\)). The performance gain of MIMO-GNN over MMSE is approximately 2.5dB at SER\(=10^{-2}\). Meanwhile, the improvement of MIMO-GNN-MMSE over MIMO-GNN is negligible. While, Fig. \ref{kro} shows the results for correlated channels with \(\rho=0.3\). MIMO-GNN maintains around 2dB gain over MMSE when SER is \(10^{-2}\). MIMO-GNN-MMSE outperforms MIMO-GNN in all the SNR range of the experiments.  Integrating the MMSE prior in the model helps to increase of 0.5dB the performance gain when SER is \(10^{-3}\).

\section{Conclusions}\label{conclusions}
We have developed MIMO-GNN, a GNN-based algorithm to solve massive MIMO detection at higher-order modulation. In contrast with BP, our experiments show that the uniform prior is sufficiently informative for MIMO-GNN to significantly outperform MMSE. This performance gain, even on correlated channels, makes MIMO-GNN a promising solution for MIMO detection. Moreover, since the computation in each iteration is done independently for every edge of the graph, the complexity of our solution can be considerably reduced by a parallelization of the algorithm.

\newpage


\bibliography{references}
\bibliographystyle{icml2020}

\end{document}